\documentclass[fleqn,twoside,twocolumn,nofootinbib,showkeys]{revtex4} 
\usepackage[sec,nocpr]{ujp} 

\begin{document}
\title[Aging of aqueous laponite dispersions]
{AGING OF AQUEOUS LAPONITE\\ DISPERSIONS IN THE PRESENCE\\ OF SODIUM POLYSTYRENE SULFONATE}%
\author{V.~Savenko}
\affiliation{Taras Shevchenko National University of Kyiv, Faculty of Physics}
\address{64/13, Volodymyrska Str., Kyiv 01601, Ukraine}
\email{savenkovolod@mail.ru}
\affiliation{Institut Charles Sadron, UPR22-CNRS, Universite de Strasbourg}%
\address{23, rue du Loess, BP 84047, 67034 Strasbourg Cedex 2,  France}%
\author{L.~Bulavin}
\affiliation{Taras Shevchenko National University of Kyiv, Faculty of Physics}
\address{64/13, Volodymyrska Str., Kyiv 01601, Ukraine}
\author{M.~Rawiso}
\affiliation{Institut Charles Sadron, UPR22-CNRS, Universite de Strasbourg}%
\address{23, rue du Loess, BP 84047, 67034 Strasbourg Cedex 2,  France}%
\author{N.~Lebovka\,}%
\affiliation{Taras Shevchenko National University of Kyiv, Faculty of Physics}
\address{64/13, Volodymyrska Str., Kyiv 01601, Ukraine}
\affiliation{F.D. Ovcharenko Institute of Biocolloidal Chemistry, Nat. Acad. of Sci. of Ukraine}%
\address{42, Blvr. Vernadskogo, Kyiv 03142, Ukraine}%
\email{lebovka@gmail.com}
\udk{538.9} 
 \razd{\secvi}


\autorcol{V.\hspace*{0.7mm}Savenko,  L.\hspace*{0.7mm}Bulavin,
M.\hspace*{0.7mm}Rawiso et al.}

\setcounter{page}{589}%

\begin{abstract}
Aqueous suspensions of Laponite with discotic particles are
well-studied and find a wide range of applications in industry.\,\,A
new direction of their implementation is polymer composites that can
exhibit improved physical properties.\,\,We have studied the aging
of aqueous suspensions of Laponite and sodium polystyrene sulfonate
(PSS--Na) and both their microscopic (small-angle X-ray scattering,
SAXS) and macroscopic (small amplitude oscillatory shear (SAOS)
rheometry) properties.\,\,The concentration of Laponite, $C_L$, was
fixed at 2.5\% wt and concentration of PSS--Na, $C_p$, was varied
within 0--0.5\% wt (0--24.2~mM).\,\,It is shown that the adding of
PSS--Na significantly accelerates the aging.\,\,Nevertheless, the
systems were stable against the sedimentation, and the flocculation
didn't occur.\,\,Polyelectrolyte induced the appearance of
large-scale fractal heterogeneities, which became more compact in
the course of the aging.
\end{abstract}


\maketitle

\section{Introduction}

Laponite is a synthetic clay, which consists  of discotic colloidal
particles with the diameter $d \approx 30$~nm and the thickness $h
\approx 1$~nm \cite{1}.\,\,In aqueous media, their faces have a
large negative charge ($\approx$$ -700$~е), while their edges are
positively charged ($\approx$$+50$~е at pH$\leq$ $\leq 11$) due to
the protonation of OH groups \cite{2}.\,\,Aque\-ous suspensions of
Laponite display a very rich phase behavior that includes sol, gel,
glass, and nematic sta-\linebreak tes \cite{3, 4, 5}.\,\,The
gelation transition was observed at a Laponite volume fraction of
$\approx0.7$\% ($\approx$$ 1.8$\% wt).\,\,It was ac\-com\-panied by
the thixotropic response to a stress~\cite{6}.

The structure of a gel is described as the spanning network of discs
connected via the so-called $T$-bonds or edge-to-face bonds
(``house-of-cards'' model) \cite{7}.\,\,Moreover, it was assumed
that the gel structure is fractal, whereas the structure of the
glassy state has a uniform density \cite{8}.\,\,Recently, the
time-dependent self-organizing processes in the so-called low-energy
states \cite{9} have obtained a considerable attention.\,\,The
various aging processes such as the phase separation (coacervation),
gelling, and glassing, which depend on the particle type, salt
concentration, and presence of adsorbed substances (surfactants and
polymers), were reported in \cite{10}.

The kinetics of a gel aging is known to depend on the  potential of
interaction between nanodisks that is a sum of the van der Waals
attraction and the electrostatic double-layer repulsion.\,\,It was
demonstrated that the addition of a monovalent salt such as NaCl
increases the dominance of the attraction between Laponite particles
in face-to-edge configurations \cite{9}.\,\,In addition, the
additional attraction or repulsion between Laponite particles can be
finely tuned, by using various additives.\,\,For example, it was
shown that the adsorption of cationic CTAB on Laponite particles
induced the additional attraction between them owing to the
enhancement of hy\-dro\-pho\-bic interactions \cite{11}, and the
adsorption of non-ionic polyethylene glycol (PEG) hindered the
gelation of Laponite owing to the steric repulsion between molecules
of the polymer adsorbed on Laponite \cite{12, 13, 14, 15, 16, 17}.
The interactions between PEG and Laponite result in the appearance
of the so-called ``shake-gel'' phenomenon even at a rather low
concentration of the polymer \cite{18, 19}.

\begin{figure}%
\vskip1mm
\includegraphics[width=0.45\textwidth]{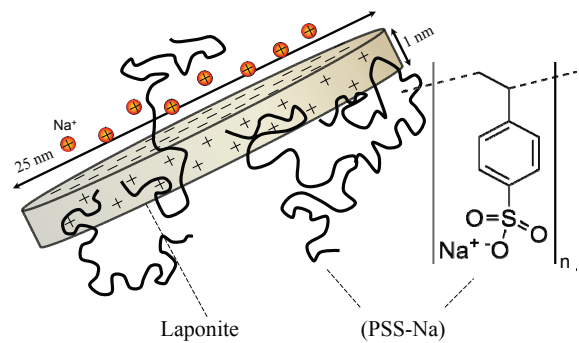}
\vskip-3mm\caption{Structure of Laponite and polystyrene sulfonic
acid-sodium salt (PSS--Na) system}
\end{figure}

For more complex pluronic tri-block copolymers, the preferential
segregation of hydrophobic segments at the surface of Laponite with
hydrophilic segments dangling into a solution was observed
\cite{20}.\,\,Elec\-tro\-sta\-tic interactions between Laponite and
gelatin-A (polyelectrolyte) was shown to drive the complex spinodal
decomposition and coacervation with the formation of ellipsoidal
domains \cite{21,22}.\,\,However, the mechanism of aging of the
electrolyte-containing Laponite gel can depend on many details of
interactions between Laponite and a polymeric electrolyte, and it is
still debated.

This work is devoted to the study of the aging of an aqueous
Laponite suspension  in the presence of PSS--Na.\,\,The
concentration of Laponite, $C_L$, in all the experiments was fixed
at $2.5$\% wt, which is just near the boundary between the agings
into the attractive gel and the repulsive ``Wigner'' glass in the
salt-free aqueous solution (below $10^{-4}$ M  concentration of
Na$^+$ ions) \cite{5}.\,\,At $C_L\approx 2.5$\% wt, the Debye
screening length in a salt-free aqueous solution is rather large, $
\lambda\approx 4.2$~nm \cite{9}, and the estimations show that the
negative potential can affect practically all the particles
\cite{16}.\,\,A glass-like state is preferably formed in a salt-free
solution, and an increase of the concentration of Na$^+$ ions leads
to the formation of a gel-like state.\,\,The concentration of
polymeric salt PSS--Na in our experiments was below $0.5$\% wt,
which corresponds to a Na$^+$ concentration of $\leq$$ 2.4\times
10^{-2}$\,M. The choice of polymeric salt PSS-Na was stipulated by
the following circumstance: PSS-Na is a strong water-soluble
polyelectrolyte with repeating chains that have both hydrophobic and
hydrophilic parts with attached negative charge (anionic chains)
\cite{23}.\,\,The electrostatic attraction between these anionic
chains and positively charged edges of Laponite particles is assumed
to be very small at small concentrations of PSS--Na.\,\,However, at
high concentrations of PSS--Na, the Na$^+$ ions effectively screen
the negative charges on Laponite particles.\,\,This enhances the
attractive interactions between the polymer and Laponite and can
result in changes in the dynamics of aging and the state of the
system.\,\,In our work, the aging processes were studied by the
methods of small amplitude oscillatory shear (SAOS) rheometry and
small-angle X-ray scattering (SAXS).

\section{Experimental}

The Laponite RD (Rockwood Additives Ltd., UK) was used
as-received.\,\,Its empirical formula is
Na$_{0.7}$[(Si$_8$Mg$_{5.5}$Li$_{0.4}$)O$_{20}$(OH)$_4$], and the
solid density is  $\approx2.53$ g/cm$^3$ \cite{24, 25}.\,\,Poly
(styrene sulfonic acid-sodium salt) (PSS--Na)
(C$_8$H$_7$SO$_3$Na)$_n$ with the molecular mass of a single chain
$m=206.2$ g/mol and the average molecular mass $M=145000$ g/mole
(i.e., $n\approx 700$) was prepared by the multiple sulfonation of
polystyrene by sulfuric acid.\,\,The sulfonation degree was close to
100\%.\,\,The obtained solution of PSS--Na was cleaned and dried.
The structure of Laponite and PSS--Na is schematically presented in
Fig.~1.

Two stock solutions of Laponite and PSS--Na were prepared by mixing
these substances with deionized ultrapure water (MilliQ) and further
ultrasonicating of mixtures using a UP 400S ultrasonic disperser
(Dr.\,\,Hielscher GmbH, Germany) at a frequency of 24~kHz and an
output power of 400~W for 15~min.\,\,The combined Laponite and
PSS--Na suspensions were prepared by mixing the stock solutions of
Laponite and PSS-Na according to the required concentration with the
further ultrasonication of the mixture for 15\,\,min.\,\,The final
concentration of Laponite in all the samples, $C_L$, was fixed at
2.5\% wt, and the concentration of PSS--Na, $C_p$, was varied within
0--0.5\% wt (0--24.2~mM).\,\,After the preparation and the cooling
to room temperature, suspensions were stirred carefully for 5 min,
and the measurements were immediately started.

A small-amplitude oscillatory shear (SAOS) rheo\-met\-ry HAAKE MARS
III (Haake, Karlsruhe, Ger\-ma\-ny) tests were done for measuring
the storage and loss $G^{\prime}$ and  $G^{\prime\prime}$ shear
moduli in the range of $10^{-1}$--$10^{3}$~Pa during the total time
of 12~h.\,\,A cone-plate fixture of 35~mm in diameter with a cone
angle of $2.0^\circ$ was used.\,\,The sample was protected from the
water evaporation during the experiment by a special cover.\,\,The
steady pre-shear at 200 s$^{-1}$ was performed for 200 s before each
test in order to homogenize the suspension before the aging
experiment \cite{16}.\,\,The dynamic time sweep test was done at the
small oscillation frequency  $ \omega = 1$~Hz, and a deformation
rate of 0.01~s$^{-1}$, which was within the linear viscoelastic
regime.\,\,The complex viscosity, $\eta^*$, was evaluated as
\begin{equation}
\eta^*=\sqrt{G^{'2}+G^{''2}}/\omega
.\label{Eq1}%
\end{equation}

The small-angle X-ray scattering spectra (SAXS) were obtained using
a digital detector Elexience, which allowed the range of
$q=(0.08\div 1.6)$~nm$^{-1}$ for the scattering vector
magnitude.\,\,The Cu--K$\alpha$ source of radiation with the
wavelength $\lambda_r =0.154$~nm was used.\,\,The ``effective''
structure factor $S(q)$ was estimated by the substitution of the
measured scattering intensity $I(q)$ to the relation
\begin{equation}
S(q)=\frac{I(q)}{K C_L M P(q)}
,\label{Eq2}%
\end{equation}
where $K=0.0445$ Mol/g$^2$ is the constant accounting for the
contrast  between Laponite and the solvent,  $M=930±190$~kg/Mol is
the molecular mass of platelets, and $P(q)$ is the form factor.
The term ``effective'' was used because relation (\ref{Eq2}) is
only valid for spherical particles.

The form factor $P(q)$ of discotic particles with diameter $d$ and
thickness $h$ was calculated using the Guinier equation
\cite{26}:\vspace*{-2mm}
\begin{equation}
p(q)=\int\limits^{\pi/2}_{0}\frac{4J^2_1(0.5qd\sin x)\sin (qh\cos
x)} {4J^2_1((qh\cos x)^2 (0.5qd\sin x)^2 }\sin x dx
,\label{Eq3}%
\end{equation}
where $J_1$ is the first-order cylindrical Bessel function.

The effect of polydispersity in the particle diameter was taken into
account  by the convolution of relation (\ref{Eq2}) with a normalized
Gaussian distribution \cite{27},
\begin{equation}
\Psi(d)=(\Delta\sqrt{\pi})^{-1}\exp\left(\!\frac{d-\langle
d\rangle}{\Delta}\!\right)^{\!\!2}\!,
\label{Eq4}%
\end{equation}
where $\langle d\rangle$ is the mean particle diameter and the
standard deviation $\Delta$ is the polydispersity parameter.\,\,The
values of $\langle d\rangle=25\pm 0.5$~nm and  $\Delta=8\pm 2$~nm
were used as the best fit parameters \cite{27}.

In the present experiments, the scattering length density (SLD)
contrast  between macroions and water was very low, and the observed
scattering patterns reflected the input of Laponite particles or
their aggregates.

The temperature was fixed at $T=298$~K in all measurements, and all
the  experiments were repeated at least three times.\,\,The Table
Curve 2D (Jandel Scientific, San Rafael, CA) software was used for
smoothing the data and for the estimation of their standard
deviations.\,\,Means and standard deviations are shown in the
figures by error bars.

\section{Results and Discussion}

Note that the transition from $G^{\prime}<G^{\prime\prime}$ to
$G^{\prime}>G^{\prime\prime}$ was observed  for all the studied
samples in the course of the aging.\,\,This behavior is typical of
the fluid-to-gel transition in suspensions of Laponite \cite{16}.
The characteristic time $t_g,$ at which
$G^{\prime}=G^{\prime\prime},$ can be defined as the time of
transition into an arrested state.

\begin{figure}%
\vskip1mm
\includegraphics[width=8cm]{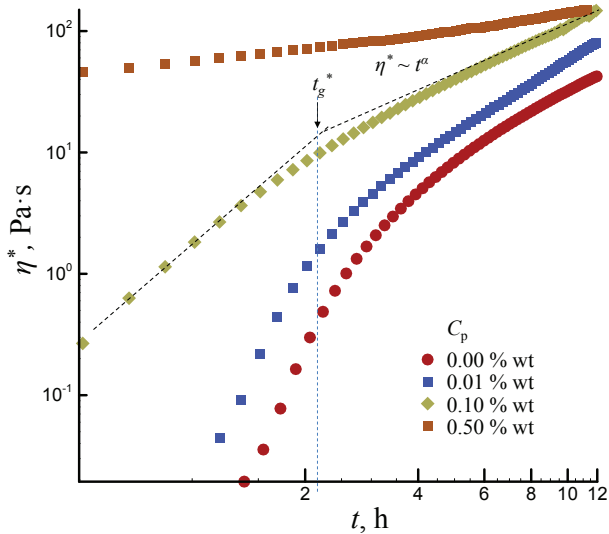}
\vskip-3mm\caption{Complex viscosity  $\eta^\ast$ versus the aging time
$t$  at various concentrations of PSS--Na, $C_p$}
\end{figure}

\begin{figure}%
\vskip3mm
\includegraphics[width=\column]{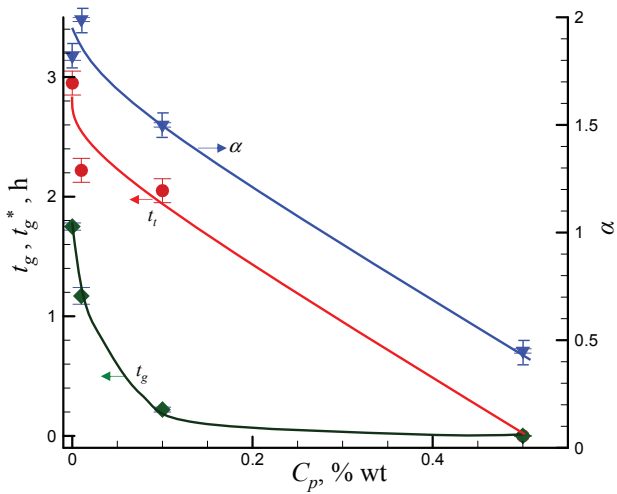}
\vskip-3mm\caption{Characteristic times, $t_g$ and $t_g^*$, and
the power-law exponent $\alpha$  (for a long time of aging) versus
the concentration of PSS--Na,~$C_p$}
\end{figure}

Figure 2 presents the time dependences of the complex viscosity,
$\eta^\ast$, during the aging at various concentrations of
polyelectrolyte $C_p$.\,\,The data show that the viscosity can
change by many orders of magnitude during the time of observation
$t=0$--12~h.\,\,The complex viscosity also shows a typical power-law
type dependence on the age \cite{8}
\begin{equation}
\eta^\ast \propto t^\alpha
,\label{Eq5}%
\end{equation}
with different slopes $\alpha$ at small and long times of aging~$t$.

The characteristic time $t_g^*,$ at which the slope changes, can be
also used for the characterization of the time of transition into an
arrested state.\,\,The corresponding dependences of $t_g$, $t_g^*,$
and the power-law exponent  $\alpha$ (for a long time of aging)
versus the polyelectrolyte concentration $C_p$ are presented in
Fig.~3.\,\,It can be seen that the introduction of PSS--Na results
in a decrease of both $t_g$, and $t_g^*,$ and the transition into
the arrested state was observed practically instantly after the
preparation of a suspension at the highest concentration of $C_p$
(=$0.5$\% wt).\,\,Moreover, the complex viscosity shows a weaker
dependence on the age with increase in $C_p$.

Note that the effect of the polyelectrolyte on the viscosity of a
solvent (i.e., at $C_L=0$\% wt) is expected to be low even at the
maximum concentration of PSS--Na, $C_p=0.5$\% wt
(=$24.2$~mM).\,\,The used range of concentrations corresponds to the
unentangled semidilute regime, when the experimental data follow the
scaling dependence $\eta\propto C_p^{0.5}.$ According to \cite{28},
the estimated level of $\eta$ should be of the order of (2--3)
$\eta_w$ (=$0.75\times 10^{-3}$ Pa$\cdot$s  at 25~$^\circ$C), where
$\eta_w$ is the viscosity of water.\,\,In general, the observed
behavior at different concentrations of PSS--Na was rather similar
to that observed in Laponite suspensions at different concentrations
of monovalent NaCl salt \cite{8}.\,\,This was explained by a
decrease of the electrostatic screening length associated with
Laponite discs with increase in the salt
concen\-tration.\looseness=1

\begin{figure}%
\vskip1mm
\includegraphics[width=8cm]{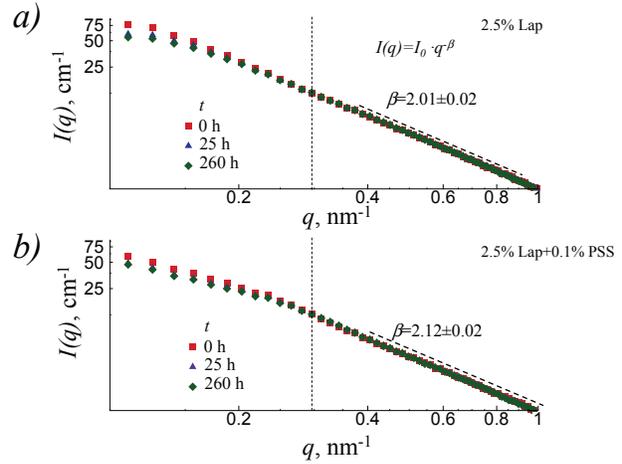}
\vskip-3mm\caption{Evolution of SAXS spectra during the aging  of
the PSS--Na free suspension (\textit{a}) and the suspension with
$C_p=$ $=0.1$\% wt (\textit{b})}
\end{figure}

Figure~4 shows the evolution of SAXS spectra during the aging  of
Laponite suspensions in the absence (\textit{a}) and in the presence
(\textit{b}) of PSS--Na.\,\,At large $q$, the decay of the scattered
intensity $I(q)$ follows roughly a $q^{-\beta}$  power-law with
$\beta=2.01 \pm 0.02$ (Fig.~4,~\textit{a}) and  $\beta=2.12 \pm0.02$
(Fig.~4,~\textit{b}).\,\,Note that the value of\mbox{ $\beta\approx
2$} is expected for randomly oriented thin disks~\cite{27}.

\begin{figure}%
\vskip1mm
\includegraphics[width=\column]{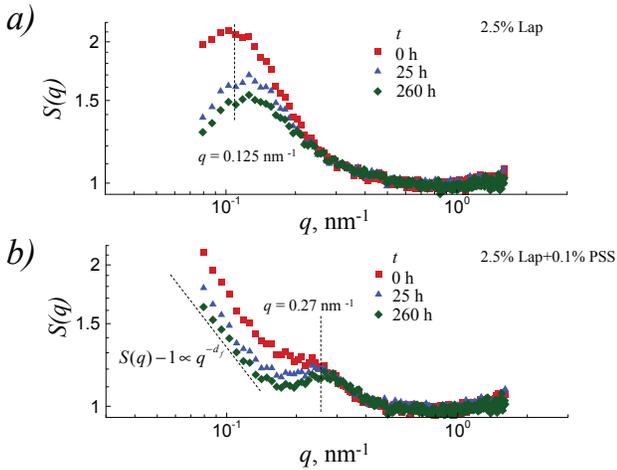}
\vskip-3mm\caption{Static structure factor $S(q)$ calculated  from
SAXS data for the pure 2.5-\% Laponite suspension (\textit{a}) and
for the Laponite suspension containing 0.1 \%wt PSS--Na
(\textit{b})}\vspace*{2mm}
\end{figure}

To extract more useful information from the scattering patterns, we
have calculated the structure factor $S(q)$ by the method as
described above.\,\,The function $S(q)$ was previously used for
determining the origin of two different non-ergodic states: gel and
Wigner glass \cite{5}.\,\,It was shown than $S(q)$ grew
significantly at small $q$-values in suspensions with Laponite
concentration under 2\% wt, which is an evidence of large-scale
(over 200~nm) heterogeneities, which are typical of the gel state.
Moreover, for samples with Laponite concentrations above 2\%, $S(q)$
was flat in the range of low $q$ values, which is the  evidence of a
homogeneous glass state.

For the pure 2.5-\% Laponite suspension, we observe a clear peak of
the structure factor $S(q)$ at $q=0.125$~nm$^{-1}$
(Fig.~5,~\textit{a}). This value obviously corresponds to the
average interparticle distance, which is about $50$ nm (two times
larger than the mean diameter of Laponite platelets).\,\,The value
of $S(q)$ was decreased continuously at low $q$ ($q<0.1$ nm).\,\,The
observed behavior was in correspondence with the formation of a
glass state in the pure 2.5-\% Laponite suspension~\cite{5}.

\begin{figure}%
\vskip1mm
\includegraphics[width=7.5cm]{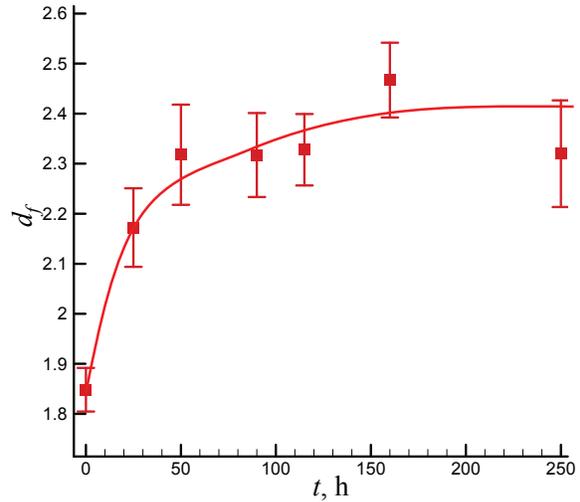}
\vskip-3mm\caption{Fractal dimension $d_f$ versus the aging time  for
the 2.5-\% Laponite suspension containing 0.1\% wt PSS--Na}
\end{figure}

However, the peak of the static structure factor of the 2.5-\%
Laponite suspension  containing 0.1\% wt PSS--Na
(Fig.~5,~\textit{b}) was shifted to larger $q$-value, if compared
with the pure Laponite suspension.\,\,Such peak matches the mean
Laponite particle diameter of 25~nm.\,\,It can evidence the presence
of T-bonds in Laponite aggregates.\,\,Moreover, a significant
increase of $S(q)$ was observed in the range of low
$q$-values.\,\,It indicates large-scale inhomogeneities, which are
typical of a gel network.

\begin{figure}%
\vskip3mm
\includegraphics[width=\column]{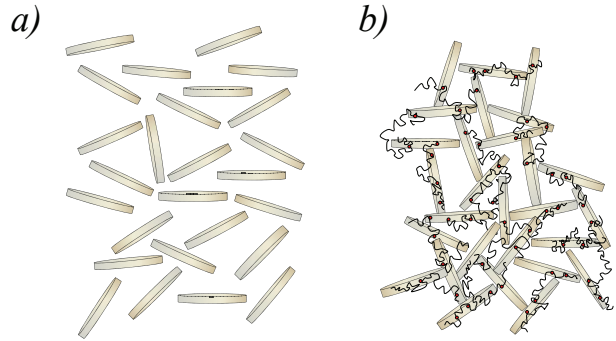}
\vskip-3mm\caption{Schemes of a glass state in the pure 2.5-\% Laponite
suspension(\textit{a}) and a fractal gel network in the 2.5-\% Laponite
suspension in the presence of 0.1\% wt PSS--Na}
\end{figure}

We used the power-law approximation
\begin{equation}
S(q)-1 \propto q^{-d_f}
,\label{Eq5}%
\end{equation}
to determine the fractal dimension of a gel structure $d_f$ in
polyelectrolyte-containing samples with various aging times
(Fig.~6).\,\,The picture shows that value of $d_f$ grows with the
aging time $t$, which means that the gel structure becomes more
homogeneous and stabilizes at $t\geq 150$ hours.

The obtained results allow to conclude that the addition of a
polyelectrolyte induces the transition in the 2.5-\% Laponite
suspension from originally glass state to gel (Fig.\,\,7), which is
typical of suspensions with lower clay concentrations ($C_L<2$\%).
This effect can be caused by a decrease of the Debye radius and thus
a partial screening of the strong electrostatic repulsion between
Laponite particles.\,\,However, the high ionic strength didn't
induce the flocculation as in the case of the doping by a monovalent
salt and samples remained stable to the sedimentation during at
least one month.\,\,The cross-linking of Laponite platelets by
polymer chains, resulting in the formation of dense fractal
aggregates by clay particles, can be the other cause of the gel
network structure \mbox{origin.}\looseness=1

\section{Conclusions}

This paper presents the results of studies of the aging in aqueous
Laponite-PSS-Na suspensions.\,\,The concentration of Laponite,
$C_L$, was fixed at 2.5\% wt, and the concentration of PSS--Na,
$C_p$, was varied within 0--0.5\% wt (0--24.2~mM).\,\,The samples
remained stable to the sedimentation and transparent at all PSS-Na
concentrations.\,\,Both visual observations and rheology tests have
shown that sodium polystyrene sul\-fo\-na\-te decreases the
ergodicity-breaking time of La\-po\-ni\-te suspensions by the factor
of 2 at a PSS--Na concentration of $0.01$\% and by the factor of 10
at a PSS concentration of $0.1$\%.\,\,The flocculation didn't appear
even at the PSS--Na concentration $C_p=0.5$\%, which corresponds to
an ionic strength of 24.2 mM, in contrast to Laponite suspensions
doped by a monovalent salt, where the flocculation was starting
immediately even at an ionic strength of 20~mM.\,\,The mechanism of
acceleration of the aggregation may reflect not only a decrease of
the Debye length, but also the cross-linking of Laponite particles
by PSS${^+}$ macroions.\,\,SAXS experiments have shown that the
addition of a polyelectrolyte induced the transition in the
2.5-\%-Laponite suspension from a homogeneous glass state to a gel
network with a fractal structure, and the fractal dimension $d_f$
increased during \mbox{the aging.}

\vskip3mm \textit{V.\,\,Savenko would like to acknowledge the
support of the Institut Charles Sadron, National Centre of Sci.
Research of France and Ministry of Edu\-ca\-ti\-on and Science of
Ukraine (grant No.\,014/60-SP).\,\,The work was sponsored in part by
the National Academy of Sciences of Ukraine (grants Nos.\,2.16.1.4,
65/13-Н and OKE/10-13).\,\,The \mbox{authors} also thank
Dr.\,N.S.\,Pi\-vo\-va\-ro\-va for her help with the manuscript
pre\-paration.}

\vspace*{-5mm} \rezume{%
В.\,Савенко, Л.\,Булавін,\\ М.\,Равізо, М.\,Лебовка} {СТАРІННЯ ВОДНИХ\\
СУСПЕНЗІЙ ЛАПОНІТУ В ПРИСУТНОСТІ\\ ПОЛІСТИРОЛ СУЛЬФОНАТУ НАТРІЮ}
{Водні суспензії лапоніту, що складаються з дископодібних частинок,
добре вивчені і знаходять широке застосування в промисловості. Новим
напрямком їх застосування є полімерні композити, які можуть мати
поліпшені фізичні властивості. В даній роботі вивчено процеси
старіння водних суспензій лапоніту в присутності полістирол
сульфонату натрію (PSS--Na). Дослідження проведено на
мікроскопічному (малокутове розсіяння рентгенівських променів) і
макроскопічному (малоамплітудна зсувна реометрія) рівнях.
Концентрація лапоніту була фіксованою, \mbox{$C_L=$}\linebreak
=~2,5\%~ваг. а концентрація PSS--Na, $C_p$, варіювалася в інтервалі
0--0,5\%~ваг. (0--24,2~ммоль). Було показано, що додавання PSS--Na
істотно прискорює старіння. Проте, вивчені системи були
седиментаційно стійкими і флокуляція не спостерігалася. Крім того, в
присутності поліелектроліту виникали великомасштабні фрактальні
неоднорідності, які ставали більш гомогенними в процесі старіння.}

\end{document}